\documentclass[aip,pop,reprint]{revtex4-1}
\usepackage{graphicx}               
\usepackage{amsmath, amssymb}       
\usepackage[T1]{fontenc}            
\usepackage[utf8]{inputenc}                              
\usepackage[ justification=justified,format=plain]{caption}
\usepackage[ justification=justified,format=plain]{subcaption}
\usepackage{color}
\usepackage[capitalise]{cleveref}
\usepackage{physics}

\begin{document}

\title{Effects of the Internal Transport Barrier in the plasma confinement with Resonant Magnetic Perturbations}

\author{P. Haerter}
\email{haerter@usp.br}
\affiliation{Instituto de Física, Universidade de São Paulo, São Paulo, São Paulo, 05508-090, Brazil}
\author{L. C. Souza }
\affiliation{Instituto de Física, Universidade de São Paulo, São Paulo, São Paulo, 05508-090, Brazil}
\author{R. L. Viana}%
\affiliation{Departamento de Física, Universidade Federal do Paraná, Curitiba, 81531-990, Paraná, Brazil.}
\affiliation{Centro Interdisciplinar de Ciência, Tecnologia e Inovação, Núcleo de Modelagem e Computação Científica, Universidade Federal do Paraná, Curitiba, 81531-990, Paraná, Brazil}
\author{I. L. Caldas}
\affiliation{Instituto de Física, Universidade de São Paulo, São Paulo, São Paulo, 05508-090, Brazil}

\date{\today}

\begin{abstract}
In the pursuit of steady-state fusion energy, Advanced Tokamak regimes rely on Internal Transport Barriers (ITBs) with reversed magnetic shear. While Resonant Magnetic Perturbations (RMPs) are commonly applied to control Edge Localized Modes (ELMs) by intentionally making the plasma edge chaotic, the RMPs can also affect the reversed-shear cores. To address this compatibility, we propose a new, adjustable analytical model for plasma current that explicitly incorporates the core, ITB, and edge pedestal components. By applying this current formulation to a Hamiltonian map, we compare how easily magnetic chaos reaches the core in ITB-driven reversed-shear (Non-Twist) plasmas. We find that the localized ITB current creates "twin" inner resonances and pushes outer magnetic island chains closer to the edge.  These results highlight a critical operational trade-off: while the Non-Twist topology makes ELM control easier at lower coil currents, it severely shrinks the safety margin against a complete, RMP-induced loss of global confinement.
\end{abstract}

\maketitle

\section{Introduction}

The pursuit of magnetically confined fusion energy relies heavily on the High-confinement mode (H-mode) operational regime~\cite{Doyle_2007,Wagner_1982,Maslov_2020}. H-mode plasmas are characterized by the spontaneous formation of an edge transport barrier (ETB)~\cite{Wagner_1984}. The suppressed turbulent transport within this barrier leads to the development of the so-called pedestal~\cite{Urano_2014,Groebner_2023,Diallo_2021}, a distinct region at the plasma boundary exhibiting exceptionally steep pressure and temperature gradients. While this pedestal is essential for achieving high global performance, its steep gradients provide the free energy to drive Edge Localized Modes (ELMs)~\cite{Leonard_ELMS_2014,Wagner_1982}. The ELMs are macroscopic magnetohydrodynamic (MHD) instabilities that can expel massive bursts of particles and heat across the plasma chamber, which can lead to severe damage to the device walls~\cite{Diallo_2021}.

To mitigate this problem, Resonant Magnetic Perturbations (RMPs) are frequently employed~\cite{Evans_RMP_2004,Sun_RPM_2016,Hu_RMP_20202,SALVADOR2025114788}. RMPs utilize strategically placed external coils to introduce small 3D field perturbations into the tokamak. These fields are tuned to resonate with the rational flux surfaces near the plasma boundary, intentionally generating localized magnetic island chains~\cite{Evans_2015}. As these islands overlap according to the Chirikov criterion, they create a controlled chaotic layer at the plasma edge. This increased continuous particle transport successfully bleeds off the pressure pedestal and prevents the buildup of large ELMs.

However, the application of RMPs presents a critical operational conflict: while controlled chaoticity is required at the edge, it is imperative to prevent this chaotic layer from penetrating into the deep core~\cite{guiHamiltonianEstimationIsland2026}. Excessive RMP amplitude can destroy the stabilizing Kolmogorov-Arnold-Moser (KAM) surfaces in the plasma interior, leading to a catastrophic loss of global confinement and triggering global disruptions~\cite{mugnaineNontwistFieldLine2023}.

The resilience of the plasma core against RMP penetration is fundamentally governed by the macroscopic safety factor profile, $q(\psi)$. In standard H-mode operation, the centralized Ohmic current yields a monotonic safety factor profile characterized by strong, strictly positive magnetic shear~\cite{Stober_2007}. In the framework of Hamiltonian mechanics, this magnetic topology corresponds to a classical ``Twist'' map.

In contrast, future steady-state fusion reactors aim to operate in Advanced Tokamak (AT) scenarios~\cite{buttery2019diii}. These regimes rely on the formation of an Internal Transport Barrier (ITB) driven by localized, off-axis bootstrap currents and auxiliary heating~\cite{Chung_2018,Ding_2017,Zhang_2025}. The introduction of an ITB fundamentally alters the current profile, creating a reversed magnetic shear configuration characterized by a local minimum in $q(\psi)$ and the emergence of shearless curves.

While the stabilizing effects of reversed shear on micro-turbulence are well documented, its macroscopic resilience against externally applied RMP chaoticity requires rigorous investigation. Specifically, the introduction of an ITB creates a topological conflict: while it generates a shearless curve, the local minimum in $q(\psi)$ also introduces multiple rational resonances for a single mode (``twin'' island chains) deep within the core. It is critical to determine whether the shearless curve acts as a protective KAM shield, or if the inner resonances act as a structural vulnerability that exacerbates chaotic penetration.

This work aims to quantify this vulnerability by comparing the chaotic penetration depths and breakdown thresholds of  Non-Twist (reversed-shear) configurations under increasing RMP field strengths. By employing a Hamiltonian mapping model that incorporates analytical profiles for the Ohmic, pedestal, and ITB currents,  using a parameter scan of the  MHD and vacuum RMP responses, we systematically assess the structural integrity of the advanced core. Ultimately, we demonstrate how the reversed-shear topology fundamentally shifts the operational trade-off, increasing the edge sensitivity for efficient ELM mitigation while simultaneously elevating the risk of premature global confinement collapse.

This paper is organized as follows:~\cref{sec:Model} details the magnetic field mapping model, the safety factor formulations, and the applied perturbation profiles. \cref{sec:PS} explores the resulting phase space topologies and the formation of resonant inner island chains. \cref{sec:Escape} quantitatively analyzes the field line escape rates, extracting the exact depth of chaotic penetration as a function of the ITB strength. Finally, Section \cref{sec:Conc} presents the summary and conclusions of our findings.

\section{The Model}
\label{sec:Model}
In this section, we consider a refined safety factor profile based on recent experimental and theoretical results. The primary objective is to update existing safety factor models to accurately analyze Advanced Tokamak (AT) operational scenarios. We rely on restrictive assumptions that fail to capture the highly localized current features of modern high-confinement regimes. Here, we propose a versatile, tunable safety factor with a formulation capable of reproducing the magnetic topology of recent advanced plasma confinement experiments. Following this, the Hamiltonian field line map and the applied magnetic perturbation functions are detailed, establishing the comprehensive mathematical framework utilized to explore our results.

\subsection{Safety Factor}

To obtain a physically consistent safety factor profile $q(\psi)$ that accurately reflects high-performance tokamak regimes, we derive the magnetic topology from a superposition of the primary current density components inherent to Advanced Tokamak (AT) scenarios. We define the total parallel current density as $$J(\psi) = J_c(\psi) + J_i(\psi) + J_e(\psi)$$, where the individual terms represent the macroscopic core current, the internal transport barrier (ITB) current, and the edge pedestal current, respectively.

The baseline topology is established by the core current $J_c(\psi)$, representing the broadly distributed inductive Ohmic current that peaks at the magnetic axis. In standard L-mode operation, this centralized current alone yields a standard monotonic (Twist) safety factor profile with strong positive magnetic shear everywhere.

However, modern high-confinement (H-mode) plasmas are characterized by the spontaneous formation of an edge transport barrier, or pedestal. The extreme localized pressure gradients within this pedestal drive a strong non-inductive neoclassical bootstrap current \cite{wadeValidationNeoclassicalBootstrap2004,kohBootstrapCurrentEdge2012}. To model this edge-localized bootstrap response, we introduce $J_e(\psi)$, defined as a narrow current peak situated near the plasma boundary. 

Furthermore, to achieve advanced steady-state regimes, current drive techniques are employed to trigger an Internal Transport Barrier (ITB)~\cite{staeblerTheoryTransportHigh2018}. The sustainment of an ITB is strongly coupled to the formation of an off-axis current peak, driven by a combination of auxiliary heating (e.g., neutral beams) and localized core bootstrap fractions \cite{chuStudyMechanismITB2021}. We model this essential feature via the internal current term $J_i(\psi)$. 

The introduction of this highly localized ITB current fundamentally alters the global magnetic topology. By carefully tuning the amplitude and radial position of $J_i(\psi)$, the global safety factor $q(\psi)$ develops a local minimum. This creates a reversed magnetic shear configuration characterized by the presence of shearless curves ($dq/d\psi = 0$).


The baseline Ohmic core current is modeled as a monotonically decreasing profile of the form:
\begin{equation}
	J_c(\psi) = J_0(1-\psi)^\nu
\end{equation}
where $J_0$ and $\nu$ are constants related to the macroscopic plasma configuration. For the present study, we set both $J_0 = 1$ and $\nu = 1$ to provide a standard linear decay.

The internal and edge transport barrier currents are modeled as localized hyperbolic secant squared peaks:
\begin{equation}
	J_{i,e}(\psi) = A_{i,e}\sech^2\qty(\frac{\psi-\psi_{i,e}}{w_{i,e}})
\end{equation}
where the subscripts $i$ and $e$ refer to the internal and external components, respectively. Here, $A$ represents the current amplitude, $\psi_{i,e}$ dictates the radial location of the peak, and $w_{i,e}$ determines the width of the transport barrier.

These parameters are chosen to qualitatively reproduce the current density profiles observed in experimental H-mode and Advanced Tokamak scenarios, as well as in prior MHD simulations. The specific values defining the used configurations are summarized in \cref{tab:Current_Par}.

\begin{table}[h!]
\centering
\begin{tabular}{c c}
\hline \hline
Parameter &  ITB Parameter Scan \\
\hline
$A_i$ &  $\in [0.0, 1.5]$ \\
$\psi_i$  & $0.4$ \\
$w_i$ &  $0.1$ \\
$A_e$ &  $0.04$ \\
$\psi_e$  & $0.95$ \\
$w_e$  & $0.02$ \\  
\hline \hline
\end{tabular}
\caption{Parameters used to define the current density profiles. The internal current amplitude ($A_i$) is treated as a continuous variable.}
\label{tab:Current_Par}
\end{table}

Defining the total parallel current density as $J(\psi) = J_c(\psi) + J_i(\psi) + J_e(\psi)$, we can integrate this profile over the radial cross-section to find the total enclosed current $I(\psi)$ up to a specific flux surface:
\begin{widetext}
\begin{equation}
	I(\psi)=\frac{J_{0}\left(1-\left(1-\psi\right)^{\nu+1}\right)}{\nu+1}+A_i w_i\left[\tanh\left(\frac{\psi-\psi_i}{w_i}\right)+\tanh\left(\frac{\psi_i}{w_i}\right)\right]+A_e w_e\left[\tanh\left(\frac{\psi-\psi_e}{w_e}\right)+\tanh\left(\frac{\psi_e}{w_e}\right)\right] 
\end{equation}
\end{widetext}

Assuming a cylindrical approximation, the global safety factor is then calculated as:
\begin{equation}
	q(\psi)=\frac{q_a I(1)}{I(\psi)}\psi
\end{equation}
where $q_a=5.6$ is the fixed value of the safety factor at the magnetic axis (center), and $I(1)$ is the total enclosed current at the plasma boundary ($\psi=1$). 

\begin{figure}[!ht]
	\includegraphics[scale=0.7]{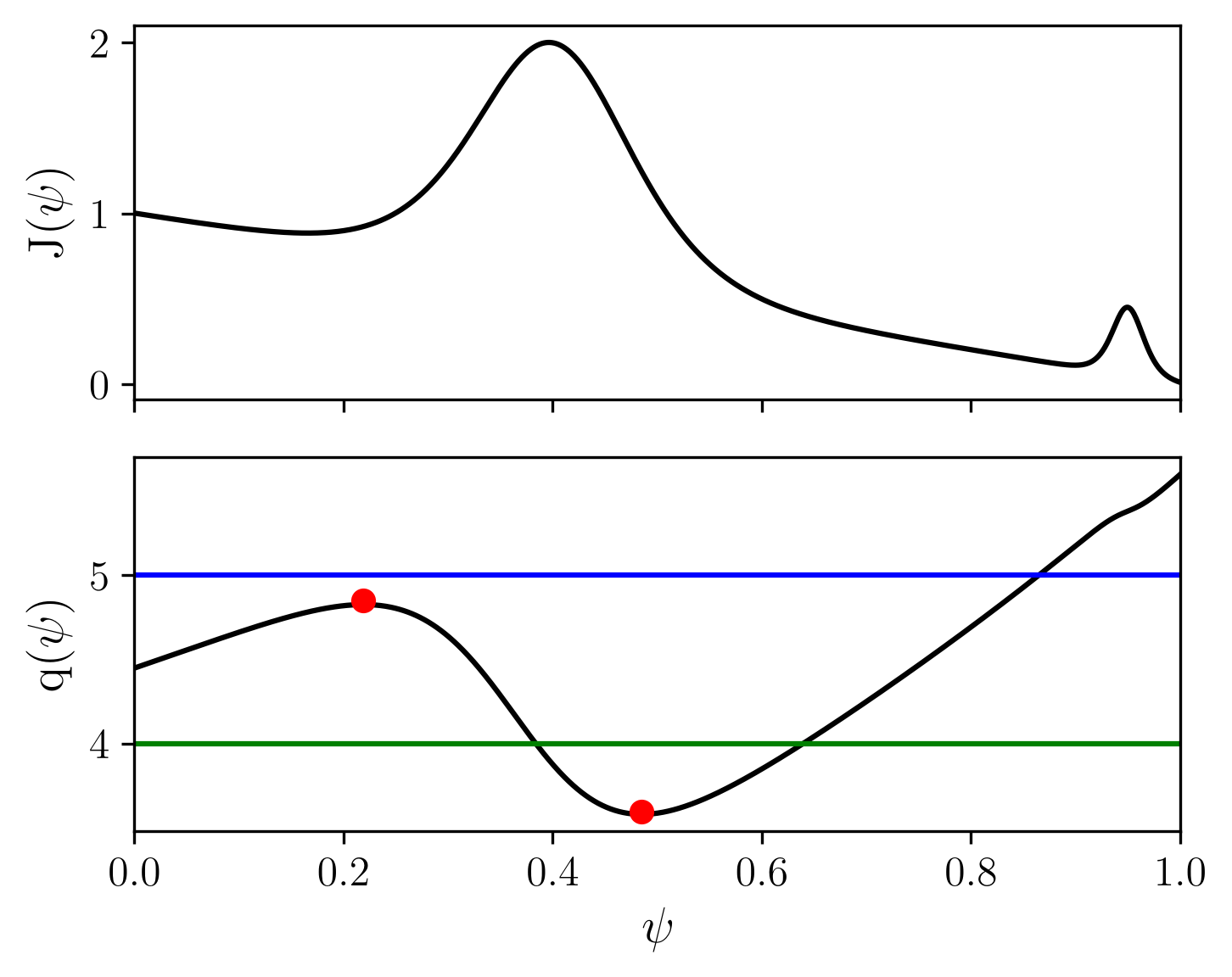}
	\caption{Plot of the safety factor and current density, with the used ressonances marked in green and blue.}
	\label{fig:Current_Factor}
\end{figure}

As illustrated in \cref{fig:Current_Factor}, this formulation allows us to explicitly compare the resulting current densities and safety factors. In the Non-Twist condition, the localized internal transport barrier induces a distinct local maximum and minimum in the $q$-profile. This creates two shearless curves that bracket the core, fundamentally altering how the plasma responds to the applied $q(\psi)=4$ and $q(\psi)=5$ resonant modes explored in this study.

\subsection{Field lines model}

The model in the present work was proposed by Abdulaev to study the field lines close to the separatrix and latter expanded to include MHD perturbations close to core\cite{guiHamiltonianEstimationIsland2026,constantinescu2011mapping}. The model is composed by,

\begin{align}
    X_n=\psi_n-\varepsilon\pdv{S}{\theta_n}(\theta_n,X_n,\phi_n)&,\quad \omega_n=\theta_n+\varepsilon\pdv{S}{X_n}(\theta_n,X_n),\\
    \bar{\omega}_n=\omega_n&+\iota(X_n),\\
    \psi_{n+1}=X_n-\varepsilon\pdv{S}{\theta_{n+1}}(\theta_{n+1},X_n,\phi_{n+1}),&\quad \theta_{n+1}=\bar{\omega}_n+\varepsilon\pdv{S}{\theta}(\theta_{n+1},X_n)
\end{align}

where $\psi,\theta$ and $\phi$ are the toroidal flux, the poloidal angle and the toridal angle. Here $\varepsilon$ is the RMP perturbation strength, and also acts as the  perturbation of the Hamiltonian system. The auxiliary variables is $\iota(X)=2\pi/q(X)$ and the generating function of the above map has the form 

\begin{equation}
\begin{aligned}
    &S(\theta,X,\phi)=\pi\sum_{m,n}H_{m,n}(\psi)\cp\\
    &\qty[a(x_{mn})\sin(m\theta-n\phi)+b(x_{mn})\cos(m\theta-n\phi)],
\end{aligned}
\end{equation}
with coefficients functions
\begin{equation}
    a(x)=\frac{1-\cos x}{x},\quad b(x)=\frac{\sin x}{x},\quad x_{mn} = m\frac{\iota(X)}{2} - n\pi,
\end{equation}
the calculation of the the derivatives of $S$ in relation to $\theta$ and $X$ are deduced in \cref{apx:Deduction}.

The amplitude profiles of the applied magnetic perturbations are governed by the functions $H_{mn}$, where $m$ and $n$ denote the poloidal and toroidal mode numbers, respectively. We adopt analytical perturbation models that represent both the internal (MHD) response of the plasma and the external resonant magnetic perturbations (RMP) to ensure the Hamiltonian map accurately reflects the toroidal geometry and boundary conditions of a tokamak.

For internal MHD activities, such as tearing modes or edge-localized modes, the perturbation must be highly localized at the resonant magnetic surface. This internal response is modeled as\cite{constantinescu2011mapping}:

\begin{equation}
    H_{mn}^{\text{MHD}}(X) = \xi\frac{n}{m} \qty[ \cosh^{-\frac{m\delta}{2}}\qty(u_1) \pm \cosh^{-\frac{m\delta}{2}}\qty(u_2) ]
\end{equation}
\begin{equation}
    \text{where} \quad u_i = \frac{1}{\delta} \ln\qty(\frac{X}{\psi_i})
\end{equation}

Here, $\psi_i$ is the radial position where the mode is resonant with the safety factor ($q(\psi_i) = m/n$), and $\delta=0.2$ is a modeled constant dictating the radial width of the mode. This specific $\cosh$ formulation is chosen because it creates a smooth peak exactly at the resonance while asymptotically reproducing the physically required cylindrical power-law decays ($X^{m/2}$ toward the core and $X^{-m/2}$ toward the edge) in the regions far from the rational surface. The parameter $\xi=0.1$ is a constant scaling factor, representing the small amplitude of the plasma's internal response relative to the external driving field.

Conversely, the external RMP field is generated by coils located outside the plasma. Therefore, its radial profile must exhibit a broad spectrum near the separatrix that decays as it penetrates inward, the simplest function which gives this is\cite{abdullaevModelMagneticField2009,abdullaevGenericMagneticField2010}:

\begin{equation}
    H_{mn}^{\text{RMP}}(X) = \frac{n}{m} X^{\frac{m}{2}} \exp\qty[-\frac{k}{2(\nu + 1)} \qty(1 - \abs{1 - X}^{\nu + 1})]
    \label{eq:H_RMP}
\end{equation}

This analytical form satisfies two critical physical constraints. First, the exponential envelope governs the radial attenuation of the externally applied vacuum fields from the plasma edge inward. Second, the $X^{m/2}$ factor enforces the strict geometric boundary condition that all non-axisymmetric multipole perturbations must vanish at the magnetic axis ($X \to 0$).

Note that the profile function in the perturbation amplitude relies strictly on the bulk unperturbed current profile. Localized variations in the current density, such as the Internal Transport Barrier and edge pedestal, are incorporated fully into the system's rotational transform $\iota(X)$ to capture exact resonance locations, but are excluded from the global perturbation decay envelope to maintain an analytical representation of the external $RMP$ fields.

To capture the distinct physical origins of the magnetic islands at different radial depths, the total perturbation amplitude for each resonant mode is constructed differently. For the outer resonance located near the plasma edge ($m=15, n=3$ at $q=5$), the rational surface is directly exposed to the external coil fields while also undergoing an internal plasma response. Therefore, the total perturbation for this mode is the superposition of the vacuum RMP envelope and the localized MHD response:

\begin{equation}
    H_{15,3}(X) = H_{15,3}^{\text{RMP}}(X) + H_{15,3}^{\text{MHD}}(X)
\end{equation}

Conversely, for the inner resonance located deep within the core ($m=12, n=3$ at $q=4$), the externally applied vacuum RMP field has decayed significantly. To isolate the effects of the internal topology, we model this deep resonance strictly through its localized internal plasma response. Thus, the total perturbation for the inner mode is defined purely by the MHD term:

\begin{equation}
    H_{12,3}(X) = H_{12,3}^{\text{MHD}}(X)
\end{equation}

For all simulations, the global scaling factor representing the perturbation strength, $\varepsilon$, is kept consistent across both modes to allow for a direct comparison of the topological breakdown thresholds. The radial locations of these $q=4$ and $q=5$ resonant surfaces are indicated by the green and blue horizontal lines, respectively, in \cref{fig:Current_Factor}.

\begin{figure*}[!ht]
    \centering
    \includegraphics[scale=0.9]{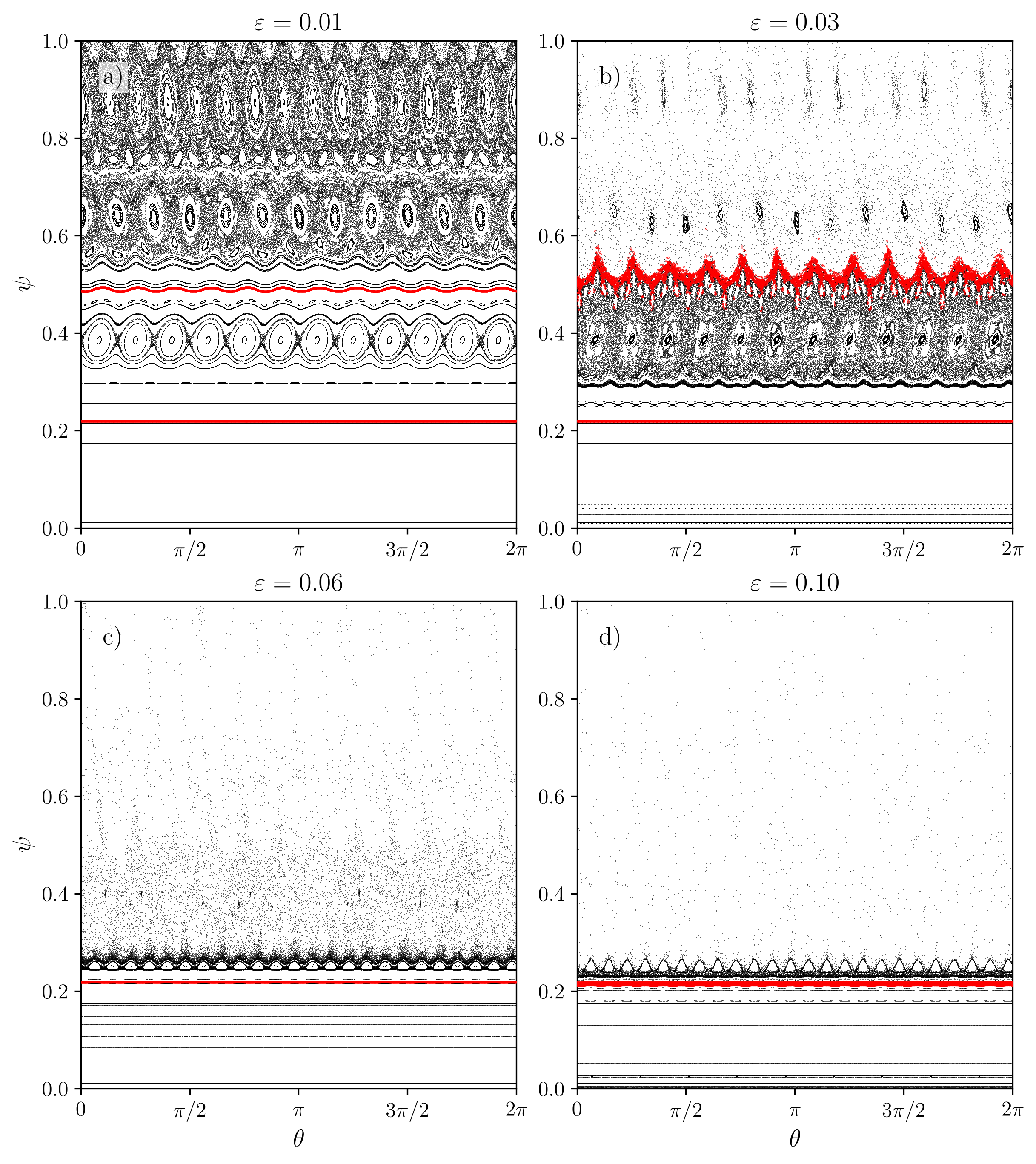}
    \caption{Poincaré phase space sections for the multiple perturbation amplitudes.}
    \label{fig:PS}
\end{figure*}

\section{Phase Space}
\label{sec:PS}

To analyze the topology of the Internal Transport Barrier (ITB) on magnetic confinement, we construct Poincaré phase space sections. A uniform grid of initial conditions was iterated through the Hamiltonian map for $N_f=10^6$ steps. By systematically varying the external RMP amplitude ($\varepsilon$), we can directly observe the dynamic structures—such as KAM tori, resonant magnetic island chains, and chaotic seas, that govern macroscopic plasma transport across three distinct perturbation regimes (\cref{fig:PS}).

Under low perturbation limit ($\varepsilon=0.01$), the baseline of the configurations is established. In presence of the ITB (~\cref{fig:PS}a)) fundamentally alters this resonant topology by driving a reversed magnetic shear configuration. 

Because the modified safety factor profile possesses a local minimum, the rational resonance conditions are satisfied at multiple radial positions. This topological shift mathematically manifests as the formation of ``twin'' $q=4$ magnetic island chains (period-12) separated by a shearless curve.

While previous models suggested that this shearless curve might act as a robust transport barrier, the intermediate perturbation regime ($\varepsilon=0.03$) reveals a critical structural vulnerability. The introduction of the ITB not only creates the inner $q=4$ resonance but also displaces the outer $q=4$ resonance radially outward, compressing the spatial gap between it and the boundary $q=5$ islands. This  perturbation amplitude, the spatial compression in the Non-Twist case proves fatal to global confinement (\cref{fig:PS}d). The outer chain of the twin $q=4$ islands expands and directly collides with the border $q=5$ island chain. The Chirikov resonance overlap criterion between these specific modes is satisfied much earlier, shattering the crucial KAM surfaces and creating a continuous chaotic pathway spanning from $\psi \approx 0.5$ to the edge.

Finally, at the highest simulated perturbation ($\varepsilon=0.06$), the system suffers severe global breakdown. However, the legacy of the early topological collapse in the reversed-shear scenario remains highly evident. The pre-existing inner resonances configuration (\cref{fig:PS}d) inadvertently bridge the chaotic edge even deeper into the plasma, pushing the chaotic boundary below $\psi \approx 0.3$. Consequently, the ITB's modification of the $q$-profile forces a premature chaotic collapse, compromising the core at strictly lower perturbation thresholds.

\section{Escape Results}
\label{sec:Escape}

Besides the macroscopic topological effects observed in the phase space, we can quantitatively measure the confinement efficiency of the magnetic configurations by analyzing the escape rate of the field lines. By integrating the Hamiltonian map, a field line is considered 'escaped' if it reaches the plasma boundary at $\psi=1$, simulating a strike on the physical wall or divertor. Assuming a field line has a length $l=\phi_n$, we measure confinement based on the number of toroidal iterations $n$ before this boundary is crossed. A grid of $2^{20}$ initial conditions was integrated to calculate the connection length of each starting coordinate. 

From this integration, two primary metrics are extracted to evaluate confinement loss. First, because the formation of the $q=5$ (period-15) island chain creates a highly chaotic edge, a field line escaping from a certain inner radius $\psi_e$ indicates that the separating KAM surfaces have broken down. By tracking the deepest radial position from which field lines escape, we can pinpoint exactly when global confinement is compromised. Second, by averaging these connection lengths over the poloidal direction $\theta$, we obtain the mean connection length as a function of the initial radial depth $\psi_0$ and the external RMP amplitude $\varepsilon$.

\begin{figure}[ht!]
    \centering
    \includegraphics[scale=0.7]{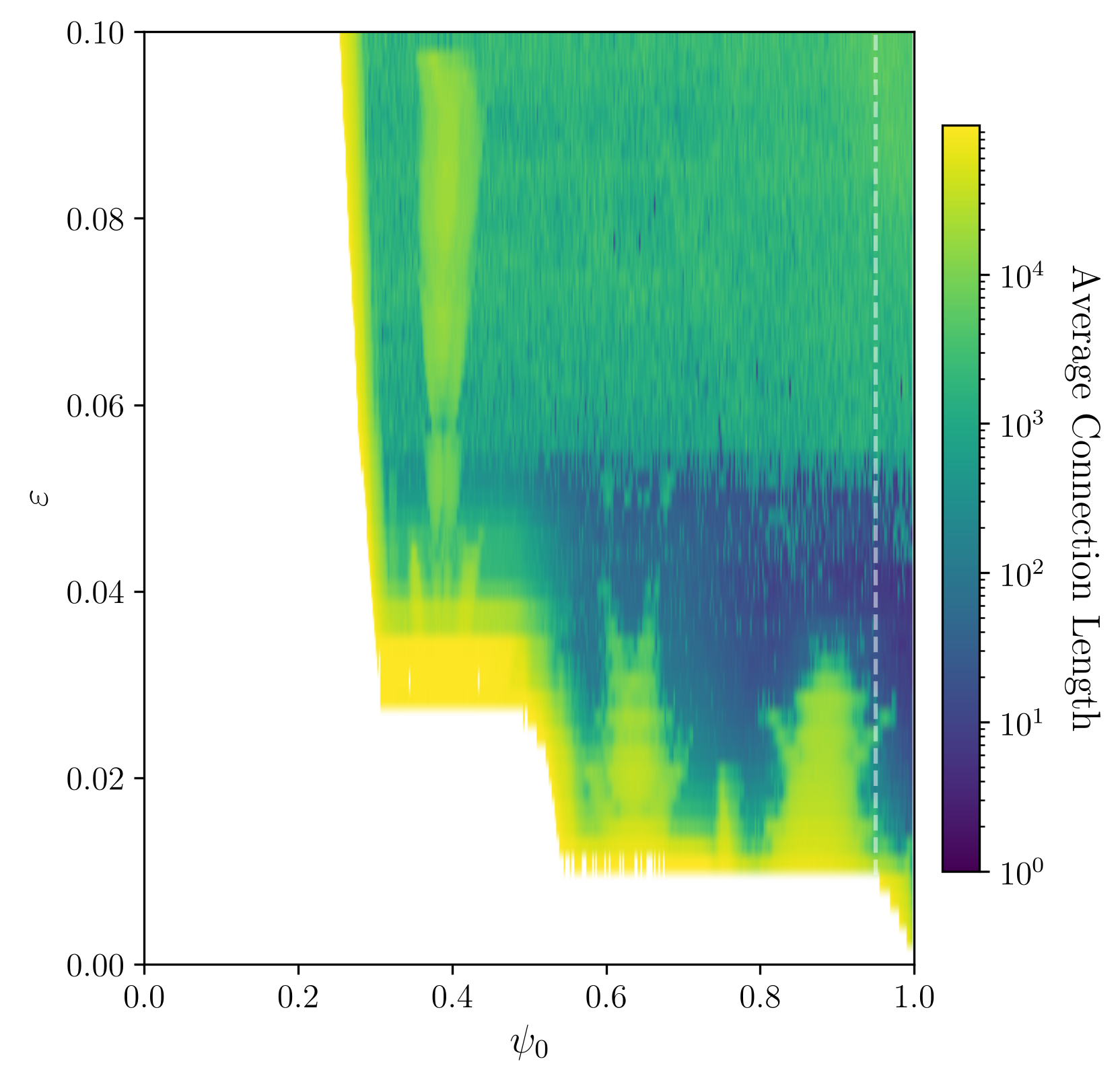}
    \caption{Average connection length mapping for multiple values of RMP amplitude $\varepsilon$ and initial radial position $\psi_0$. High values (yellow) indicate robust confinement on closed flux surfaces, while low values (blue) indicate rapid chaotic escape.}
    \label{fig:escape_time}
\end{figure}

The spatial distribution of the average connection lengths for the advanced tokamak configuration ($A_i=1.4$) is presented in \cref{fig:escape_time}. The map displays a pronounced vulnerability to core erosion. The presence of the inner resonances in the reversed-shear topology allows chaotic field lines to deeply penetrate the plasma ($\psi_0 \approx 0.6$) at low RMP amplitudes ($\varepsilon \approx 0.01$ to $0.03$). These deep structural intrusions visually confirm that the inner island chains act as a conduit, bridging the deep core to the chaotic edge.

This quantitative extraction highlights the specific vulnerabilities of the reversed-shear topology to 3D fields. Once this geometry experiences its breakdown at $\varepsilon \approx 0.01$, the resulting chaotic invasion is significantly deep. It sacrifices a substantial fraction of the burning core volume, plateauing at over $70\%$ penetration. 

However, a closer analysis of the distinct shape of the connection length distribution reveals a potentially highly advantageous feature for Edge Localized Mode (ELM) mitigation. The primary objective of RMP application is to deliberately creates chaos at the plasma edge to increase continuous particle transport, thereby preventing the steep pressure gradients that drive ELMs. The connection length heatmaps demonstrate that the reversed-shear topology generates a broad, effective chaotic edge layer (characterized by the rapid onset of the deep blue escape regions) at exceptionally low perturbation amplitudes. The radial displacement of the outer $q=4$ resonance essentially "primes" the edge for chaoticity.

These findings present an operational trade-off for Advanced Tokamak scenarios. The reversed-shear profile is highly efficient for ELM suppression, theoretically allowing operators to achieve the necessary edge chaoticity utilizing lower external coil currents, which reduces hardware stress. However, this efficiency comes at the cost of a drastically reduced safety margin. If the applied RMP amplitude slightly overshoots the narrow operational window, the pre-existing inner resonances will act as a topological bridge, triggering extreme chaotic degradation of the deep core.

\section{Chaotic region}
\label{sec:Area}

The escape analysis of \cref{sec:Escape} measures confinement through the field lines that reach the wall. This is the relevant criterion for operational tokamaks conserning the particle and heat loads, but it is blind to field lines that are fully chaotic, and wander radially over a large fraction of the minor radius, but still never reach $\psi = 1$ within the integration time. Such a line is counted as confined by the connection-length metric, yet it no longer lies on a closed flux surface and contributes to radial transport. To capture the full extent of the chaos that the RMPs generate in the reversed-shear configuration, we measure the area of the chaotic region of the Poincar\'e section directly, regardless of whether it is connected to the boundary.

Each point of the Poincar\'e section is classified with the Smaller Alignment Index (SALI)~\cite{skokosAlignmentIndicesChaos2001,skokosDetectingOrderChaos2004}, which evolves two deviation vectors $\vb{v}_1, \vb{v}_2$ along the orbit and measures
\begin{equation}
    \mathrm{SALI}(n) = \min\qty(\norm{\vb{v}_1+\vb{v}_2},\ \norm{\vb{v}_1-\vb{v}_2}),
    \label{eq:sali}
\end{equation}
with both vectors renormalised at every iteration. On a chaotic orbit the two vectors align with the most unstable direction and the index collapses exponentially at a rate set by the difference of the two largest Lyapunov exponents~\cite{RolimCSF2026}.

Given that the tangent space is two dimensional, on an invariant curve both deviation vectors are eventually squeezed onto the tangent of that curve, so the SALI decays for regular orbits as a power law in $n$ rather than exponentially \cite{skokosDetectingOrderChaos2004}. In our system a regular orbit of the core gives $\mathrm{SALI} = 9.0\times10^{-3},\ 3.6\times10^{-3},\ 1.8\times10^{-3}$ and $9.3\times10^{-4}$ at $n = 200,\ 500,\ 1000$ and $2000$, i.e.\ a clean $1/n$ decay, while chaotic orbits reach numerical underflow within about a hundred iterations. Thus a fixed threshold is therefore, to separate regular and irregular orbits is unsafe. We instead take the cut from the data: the distribution of $\log_{10}\mathrm{SALI}$ is sharply bimodal, and the separatrix is placed at the minimum between the two peaks, which for the integration times used here falls at $\approx 10^{-16}$. Field lines that reach the wall are counted as part of the chaotic region, they left through the chaotic layer. The calculations were performed with the \texttt{pynamicalsys} toolkit~\cite{salesPynamicalsysPythonToolkit2025}. In our computation, we use $N = 2500$ random initial conditions per parameter point, with a integration time of $n=10000$.

\begin{figure}[h!]
    \centering
    \includegraphics[width=\columnwidth]{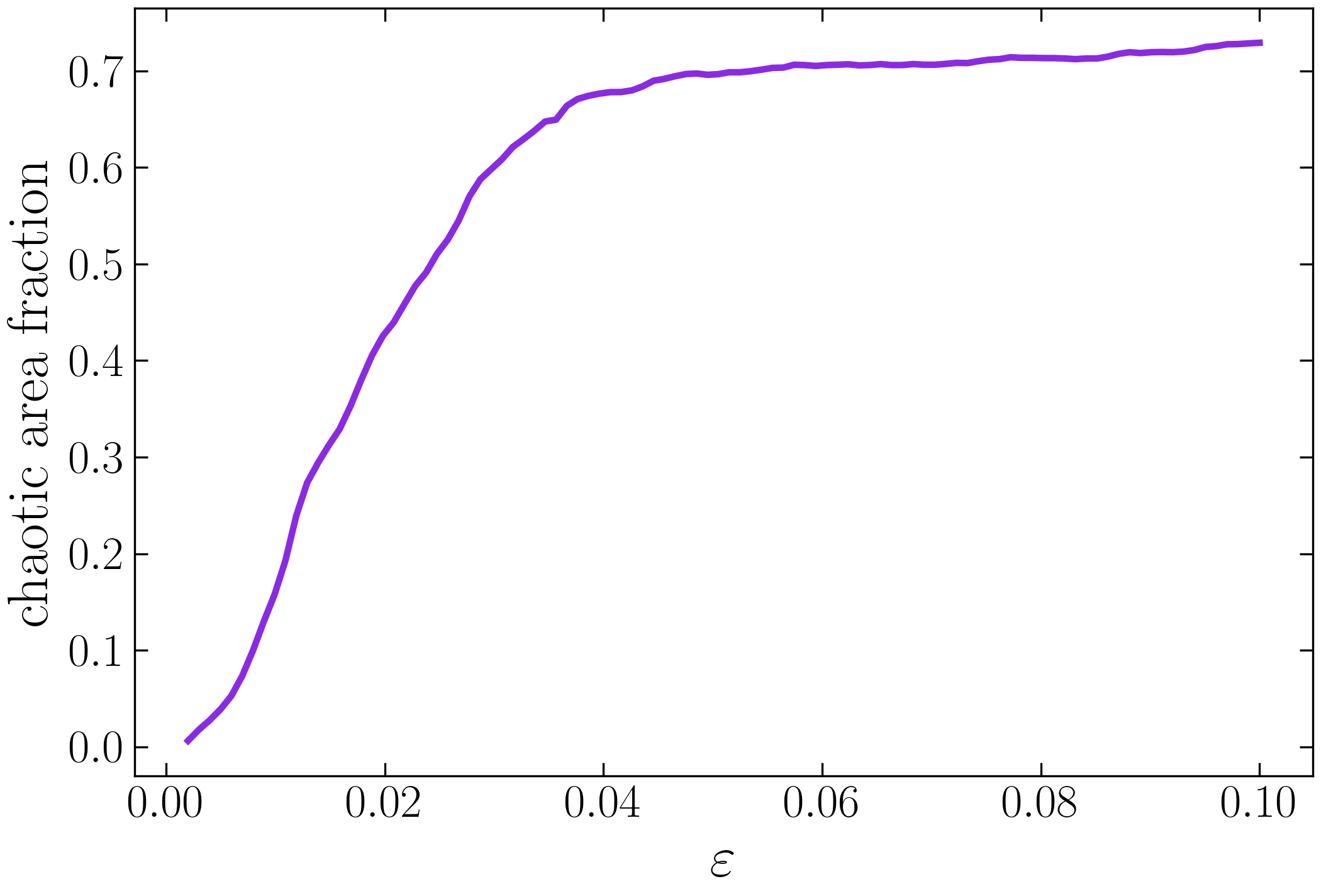}
    \caption{Chaotic area fraction of the Poincar\'e section.}
    \label{fig:area_eps}
\end{figure}

In Figure~\Cref{fig:area_eps} is show how the chaotic area of the reversed shear $(A_i=1.4)$ grows with the RMP amplitude. Chaos is present even for very small values of $\varepsilon$. The chaotic fraction is $0.04$ at $\varepsilon = 0.005$ and $0.16$ at $\varepsilon = 0.010$, the amplitude at which the escape analysis locates the loss of global confinement. It then rises steeply, to $0.43$ at $\varepsilon = 0.020$ and $0.60$ at $\varepsilon = 0.030$, and saturates beyond $\varepsilon \approx 0.04$, increasing only from $0.68$ to $0.73$ between $\varepsilon = 0.040$ and $0.100$.

These three stages follow the sequence of resonance overlaps described in \cref{sec:PS}. At the lowest amplitudes the chaos is confined to thin chaotic layers around the separatrices of the twin $q = 4$ chains and of the $q = 5$ edge chain. The steep rise starts once the islands are wide enough to overlap. Because the reversed shear displaces the outer $q = 4$ resonance towards the edge, the gap between this chain and the $q = 5$ chain is narrow, the Chirikov criterion is met early, and the KAM surfaces between the two chains are destroyed; at $\varepsilon \approx 0.03$ a continuous chaotic region already extends from $\psi \approx 0.5$ to the edge. Beyond this point the inner $q = 4$ chain acts as a bridge that carries the chaos deeper into the plasma, and at $\varepsilon \approx 0.06$ the chaotic boundary lies below $\psi \approx 0.3$. The saturation of the chaotic fraction near $0.7$ mirrors the plateau of the penetration depth found in \cref{sec:Escape}: the regular core that remains inside this boundary is eroded only slowly by a further increase of the perturbation. For ELM control this behavior cuts both ways. A modest coil current already produces a sizable chaotic region, but that region grows so quickly between $\varepsilon \approx 0.01$ and $0.03$ that a small overshoot turns an edge layer into a chaotic sea covering more than half of the section.

\begin{figure}[h!]
    \centering
    \includegraphics[width=\columnwidth]{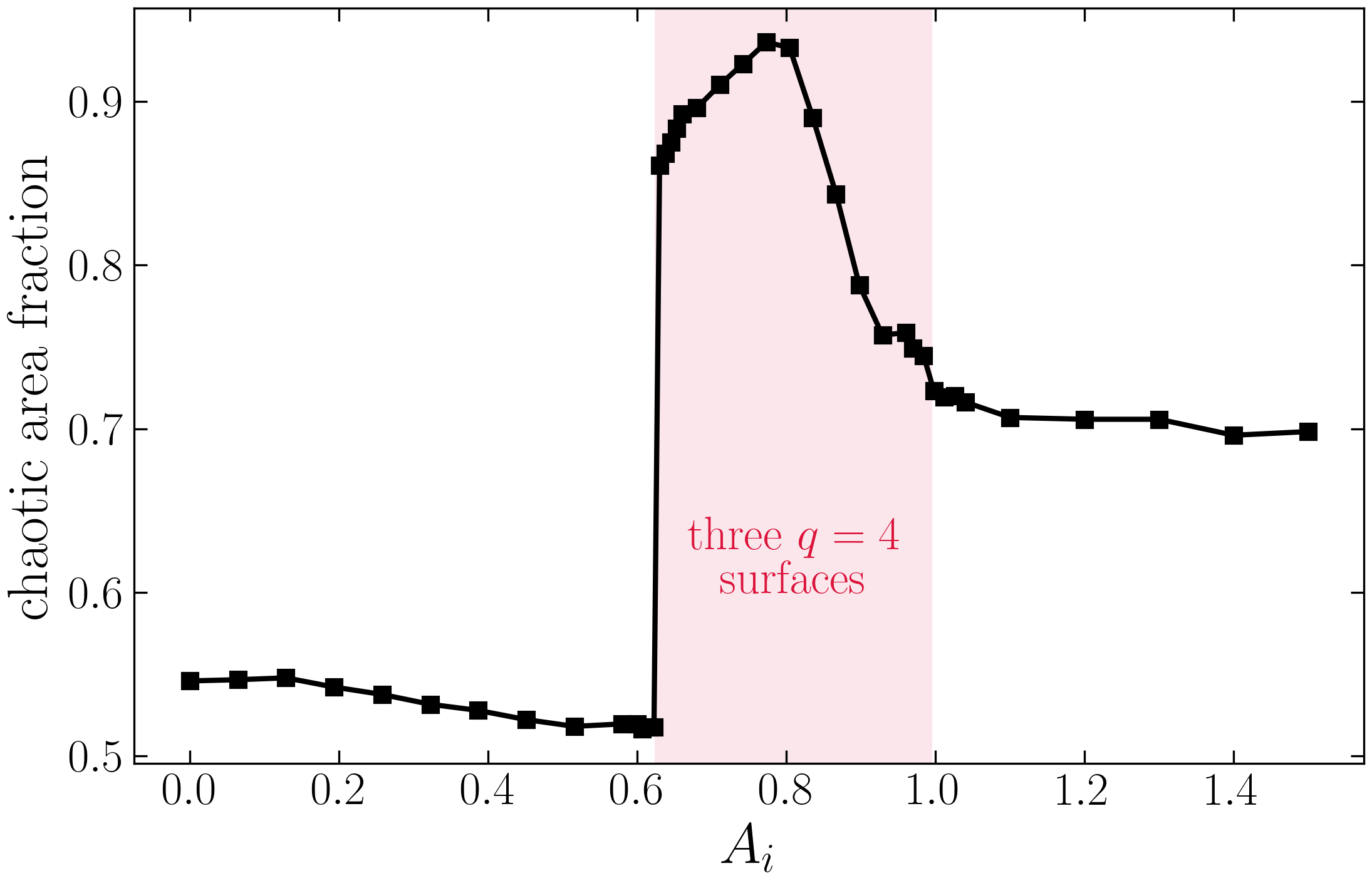}
    \caption{Chaotic area fraction against ITB current amplitude at $\varepsilon = 0.05$ the red region marks the interval in which $q(\psi) = 4$ has thre roots.}
    \label{fig:area_aitb}
\end{figure}

Scanning the ITB current amplitude reveals that the dependence on $A_i$ is not monotonic (\cref{fig:area_aitb}). At $\varepsilon = 0.05$ the chaotic area rises abruptly near $A_i \simeq 0.62$, from $0.53$ to a maximum of $0.94$, and recedes again above $A_i \simeq 1.0$ to the $0.70$ of the fully formed barrier at $A_i = 1.4$. The origin of this window can be undestood from the safety factor. As $A_i$ grows, the local minimum of $q(\psi)$ deepens; at $A_i = 0.6233$ it touches the value $q = 4$ from above, and a pair of resonant surfaces is born in a tangent bifurcation. Over the interval $0.6233 \lesssim A_i \lesssim 0.9954$, the condition $q(\psi) = 4$ is satisfied at three radii rather than two. The innermost of the three migrates towards the magnetic axis as $A_i$ increases, passing through $\psi = 0.254,\ 0.183,\ 0.116$ and $0.056$ at $A_i = 0.63,\ 0.70,\ 0.80$ and $0.90$, and leaves the plasma through the axis at the upper end of the interval. The twin chains discussed in \cref{sec:PS} for $A_i = 1.4$ are thus the survivors of a triplet. Three chains of the same poloidal mode overlap far more readily than two~\cite{chirikovUniversalInstability1979}, and throughout the window the chaotic region floods the plasma column.

\begin{figure}[ht!]
    \centering
    \includegraphics[width=\columnwidth]{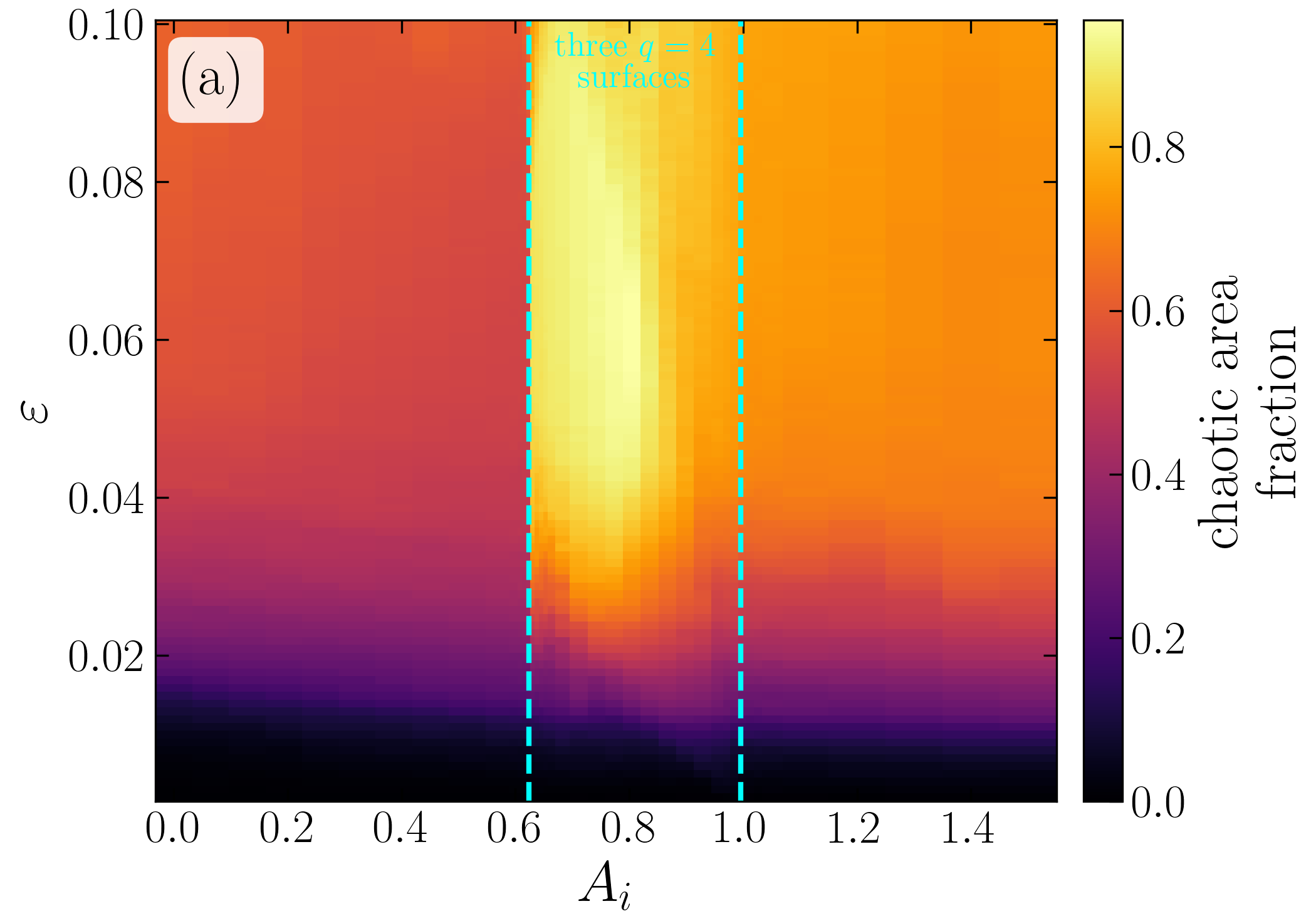}
        \includegraphics[width=\columnwidth]{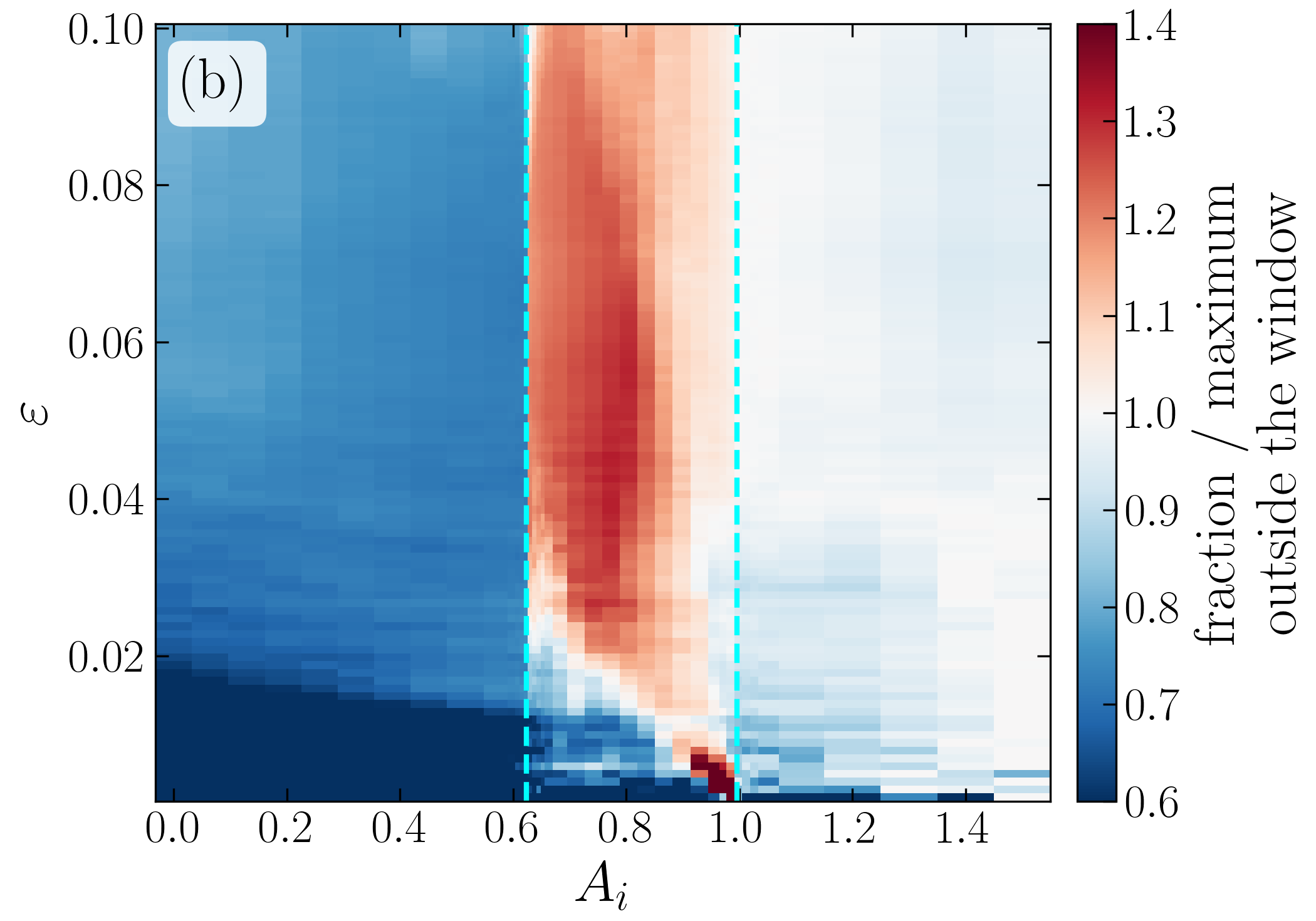}
    \caption{Chaotic area fraction over the $(\varepsilon, A_i)$. In (a) absolute fraction. In (b) the same normalised to the largest value at the same $\varepsilon$ outside the window, which isolates the enhancement due to the third resonance. Dashed lines are the analytic window boundaries $A_i = 0.6234$ and $0.9953$.}
    \label{fig:plane}
\end{figure}

Mapping the chaotic fraction over the full $(\varepsilon, A_i)$ plane (\cref{fig:plane}) shows that the window, of increase of chaos, switches on and moves. It has a threshold in perturbation amplitude: below $\varepsilon \simeq 0.015$ the chaotic area is nearly independent of $A_i$, because three chains are present as soon as $q(\psi)=4$ has three roots but overlap only once they are wide enough. The enhancement relative to the same $\varepsilon$ outside the window then grows to a maximum of $1.31$ at $\varepsilon \simeq 0.043$ and declines again at larger amplitude, where the section is close to fully chaotic in either configuration. Theofore, our result show that the most damaging amplitude is itself a function of $\varepsilon$. The peak enters at the upper edge of the window, $A_i \simeq 0.87$ at $\varepsilon = 0.013$, and migrates steadily inward with increasing perturbation, reaching $A_i \simeq 0.77$ over $\varepsilon = 0.03$--$0.08$ and $A_i \simeq 0.68$ at $\varepsilon = 0.10$. There is therefore no single worst ITB amplitude to avoid the hazardous region is a tilted band in the $(\varepsilon, A_i)$ plane, and an operating point judged safe at one coil current may not be at another.

\section{Conclusion}
\label{sec:Conc}
This study evaluated the macroscopic resilience of Advanced Tokamak plasmas against externally applied Resonant Magnetic Perturbations (RMPs). Using a tunable analytical model for plasma current, we demonstrated that the localized off-axis Internal Transport Barrier (ITB) current fundamentally alters the magnetic structure in reversed-shear configurations by creating "twin" inner rational resonances and displacing outer magnetic islands toward the chaotic edge.

Because of this compressed topological layout, the reversed-shear profile proved highly vulnerable to external perturbations. The geometry suffered a premature chaotic collapse, with global confinement breaking down at roughly $\varepsilon \approx 0.01$. Once the protective magnetic surfaces were destroyed, the pre-existing inner resonances acted as a topological bridge, pulling chaotic field lines deep into the plasma and sacrificing over 70\% of the minor radius. 

A complementary measurement of the chaotic area, obtained from the SALI classification, refines this picture in two respects. First, the fully formed ITB generates a massive expansion of the chaotic area at $\varepsilon = 0.010$, the regime in which ELM control would be exercised. Second, and unexpectedly, the dependence on the ITB current amplitude is not monotonic. Between $A_i = 0.6234$ and $A_i = 0.9953$, the local minimum of the safety factor is deep enough that the condition $q(\psi) = 4$ is met at three radii rather than two, and the resulting triplet of island chains overlaps far more readily than the twin chains of the fully formed barrier. Within this window, the chaotic region occupies up to $94\%$ of the section and reaches the magnetic axis. Moreover, the region where this effect occurs is a tilted band in the $(\varepsilon, A_i)$ plane. 

Ultimately, these findings expose a major operational trade-off for future steady-state fusion reactors. The reversed-shear profile efficiently "primes" the edge for chaoticity, allowing for effective ELM mitigation using lower external coil currents. However, this efficiency comes at the cost of a drastically reduced safety margin. Operators of Advanced Tokamak scenarios must carefully manage this narrow window, as even a slight overshoot in the applied 3D fields risks connecting the chaotic edge to the inner core, triggering a catastrophic loss of global confinement.

\section*{Acknowledgments}

P. Haerter, L. C. Souza, and I. L. Caldas acknowledge the São Paulo Research Foundation (FAPESP), Brasil, under Process Numbers 2025/28656-4, 2023/16146-6, and 2024/05700-5, respectively. R. L. Viana was supported by CNPq (Brazil) under Grants No. 403120/2021-7 and No. 301019/2019-3. This research was also supported by INCT-NeuroComp (CNPq Grant No. 408389/2024-9).

\appendix

\section{Derivatives of the Generating Function}
\label{apx:Deduction}

To integrate the Hamiltonian mapping model for the magnetic field lines, it is necessary to compute the explicit partial derivatives of the generating function $S(\theta,X,\phi)$. The derivatives with respect to the poloidal angle $\theta$ and the action variable $X$ are given by:

\begin{widetext}
    \begin{align}
    \pdv{S}{\theta} &= \pi \sum_{m,n} A_{m,n}(X) m \qty[ a(x_{mn}) \cos(\Phi_{mn}) - b(x_{mn}) \sin(\Phi_{mn}) ], \\
    \pdv{S}{X} &= \pi \sum_{m,n} \left\{ \dv{A_{m,n}}{X} \qty[ a(x_{mn}) \sin(\Phi_{mn}) + b(x_{mn}) \cos(\Phi_{mn}) ] \right. \nonumber \\
    &\qquad\qquad \left. + A_{m,n}(X) \dv{x_{mn}}{X} \qty[ a'(x_{mn}) \sin(\Phi_{mn}) + b'(x_{mn}) \cos(\Phi_{mn}) ] \right\},
\end{align}
\end{widetext}

Computing the radial derivative $\pdv*{S}{X}$ requires the application of the chain rule to the resonance argument $x_{mn}$ and the total perturbation amplitude $A_{m,n}(X)$. The phase $\Phi_{mn}$, the argument $x_{mn}$, and its radial derivative are defined as:

\begin{equation}
    \Phi_{mn} = m\theta - n\phi, \quad x_{mn} = m\frac{\iota(X)}{2} - n\pi, \quad \dv{x_{mn}}{X} = \frac{m}{2} \dv{\iota}{X},
\end{equation}

The total amplitude of the applied perturbation, $A_{mn}(X)$, depends on the specific resonant mode. It is constructed as a piecewise function representing either an internal magnetohydrodynamic (MHD) response (Type 0) or a superposition of the internal response with an external Resonant Magnetic Perturbation (RMP) field (Type 1):

\begin{equation}
    A_{m,n}(X) = 
    \begin{cases} 
        0.1 H_{\text{MHD}}(X),  \text{(Type 0)} \\ 
        H_{\text{RMP}}(X) + 0.1 H_{\text{MHD}}(X), & \text{(Type 1)} 
    \end{cases} 
\end{equation}

Consequently, the spatial derivative of the perturbation amplitude is:

    \begin{equation}
        \dv{A_{m,n}}{X} = 
    \begin{cases} 
        0.1 \dv{H_{\text{MHD}}}{X},  \text{(Type 0)} \\ 
        \dv{H_{\text{RMP}}}{X} + 0.1 \dv{H_{\text{MHD}}}{X}, & \text{(Type 1)} 
    \end{cases}
    \end{equation}

The derivative of the resonance argument $x_{mn}$ is directly dependent on the global magnetic shear. This requires evaluating the radial derivatives of the total enclosed current $I_{\text{enc}}(X)$: 

\begin{align}
    \dv{I_{\text{enc}}}{X} &= J_0 \abs{1 - X}^\nu \operatorname{sgn}(1 - X) +\nonumber \\ 
    &\frac{A_{\text{itb}}}{\cosh^2\qty(\frac{X - \psi_{\text{itb}}}{w_{\text{itb}}})} + \frac{A_{\text{ped}}}{\cosh^2\qty(\frac{X - \psi_{\text{ped}}}{w_{\text{ped}}})},
\end{align}

With the macroscopic safety factor $q(X)$, and the rotation trasnform $\iota(X)$
\begin{align}
\dv{q}{X} &= q(X) \qty[ \frac{1}{X} - \frac{1}{I_{\text{enc}}(X)} \dv{I_{\text{enc}}}{X} ], \\
    \dv{\iota}{X} &= -\frac{2\pi}{q^2(X)} \dv{q}{X},
\end{align}

Finally, to close the system, we calculate the explicit spatial derivatives of the analytical perturbation envelope functions $H_{MHD}(X)$ and $H_{RMP}(X):$

\begin{widetext}
\begin{align}
   \dv{H_{\text{MHD}}}{X} &= -\frac{n}{2X} \qty[ \cosh^{-\qty(\frac{m\delta}{2} + 1)}\qty(u_1) \sinh(u_1)  \pm \cosh^{-\qty(\frac{m\delta}{2} + 1)}\qty(u_2) \sinh(u_2) ], \\
    \dv{H_{\text{RMP}}}{X} &= H_{\text{RMP}}(X) \qty[ \frac{m}{2X} - \frac{k}{2} \abs{1 - X}^\nu \operatorname{sgn}(1 - X) ],
\end{align}
\end{widetext}

\bibliography{Refs.bib}
\end{document}